\documentclass[12pt, letterpaper]{extarticle}

\usepackage{hyperref}
\usepackage{makecell}
\usepackage{url}
\usepackage{adjustbox}
\usepackage[flushleft]{threeparttable}
\usepackage{booktabs}
\usepackage{graphics}
\usepackage{helvet}
\usepackage{times}
\usepackage{comment}
\usepackage{placeins}
\usepackage{multirow}
\usepackage{rotating}
\usepackage{mathtools}
\usepackage{subfig}
\usepackage{arydshln}
\usepackage{float}
\usepackage{multicol}
\usepackage{supertabular}
\usepackage{booktabs}
\usepackage[framemethod=TikZ]{mdframed}
\usepackage{fullpage}
\usepackage[switch]{lineno}
\usepackage{amsmath}
\usepackage{amssymb}
\usepackage{rotating}
\usepackage{array}
\usepackage{mathtools}
\usepackage[ruled]{algorithm2e}
\usepackage{algorithmic}
\usepackage{bm}
\usepackage{breqn}
\usepackage{comment}
\usepackage{enumitem}
\usepackage{graphics}
\usepackage{graphicx}
\usepackage{latexsym}
\usepackage{mathrsfs}
\usepackage{morefloats}
\usepackage{todonotes}
\usepackage{nicefrac}
\usepackage{authblk}
\usepackage{ulem}

\definecolor{normgreen}{rgb}{0.0, 0.5, 0.0}

\title{Gender inequality and self-publication patterns among scientific editors}

\author[1]{Fengyuan M. Liu}
\author[2,3]{Petter Holme}
\author[4,5]{Matteo Chiesa}
\author[6*]{Bedoor AlShebli}
\author[1*]{Talal Rahwan}

\affil[1]{\normalsize
Computer Science, Science Division, New York University Abu Dhabi, UAE.}

\affil[2]{\normalsize Department of Computer Science, Aalto University, 02150 Espoo, Finland}

\affil[3]{\normalsize Center for Computational Social Science, Kobe University, Kobe 657-8501, Japan}

\affil[4]{\normalsize Khalifa University of Science and Technology, UAE.}

\affil[5]{\normalsize UiT – The Arctic University of Norway, Norway.}

\affil[6]{\normalsize Social Science Division, New York University Abu Dhabi, UAE.}

\affil[*]{\footnotesize Joint corresponding authors. E-mails:\ bedoor@nyu.edu;\ talal.rahwan@nyu.edu}

\date{}

\nolinenumbers

\begin{document}
\maketitle

\subsection*{Abstract}
Academic publishing is the principal medium of documenting and disseminating scientific discoveries. At the heart of its daily operations are the editorial boards. Despite their activities and recruitment often being opaque to outside observers, they play a crucial role in promoting fair evaluations and gender parity. Literature on gender inequality lacks the connection between women as editors and as research-active scientists, thereby missing the comparison between the gender balances in these two academic roles. Literature on editorial fairness similarly lacks longitudinal studies on the conflicts of interest arising from editors being research active, which motivates them to expedite the publication of their papers. We fill these gaps using a dataset of 103,000 editors, 240 million authors, and 220 million publications spanning five decades and 15 disciplines. This unique dataset allows us to compare the proportion of female editors to that of female scientists in any given year or discipline. Although women are already underrepresented in science (26\%), they are even more so among editors (14\%) and editors-in-chief (8\%); the lack of women with long-enough publishing careers explains the gender gap among editors, but not editors-in-chief, suggesting that other factors may be at play. Our dataset also allows us to study the self-publication patterns of editors, revealing that 8\% of them double the rate at which they publish in their own journal soon after the editorship starts, and this behavior is accentuated in journals where the editors-in-chief self-publish excessively. Finally, men are more likely to engage in this behaviour than women.

\subsection*{Significance Statement}
Although editors play a key role in academia, the literature lacks longitudinal studies quantifying both the gender inequality among editors, as well as the rate at which editors publish in their own journal. To fill this gap, we analyze 103,000 editors across 15 disciplines spanning five decades. Although women are already underrepresented among scientists (26\%), they are even more so among editors (14\%) and editors-in-chief (8\%). The lack of women with long-enough publishing careers explains the gender gap among editors, but not editors-in-chief. We also find that 8\% of editors double the rate at which they publish in their own journal soon after the editorship starts; this behavior is accentuated among men, and among journals where the editors-in-chief self-publish excessively.

\section*{Introduction}

Editors play a central role in the process of generating scholarly knowledge. By having the final say about what gets published~\cite{newton2010quality}, they exert considerable control on the scientific discourse, thus acting as ``opinion formers, gatekeepers and arbiters of disciplinary values'' \cite{burgess2010editorial}. Some editors are full-time professionals, e.g., those handling journals such as \textit{Cell}, \textit{Nature}, and \textit{Science}. In contrast, for a great majority of journals, editors are research-active academics who perform editorial duties as part of their community service, i.e., unpaid work for the benefit of their academic discipline. For the latter type of editors, the double loyalties they hold---to the standards of science and to their own research career---create a conflict of interest that may tempt editors to exploit their power by facilitating the publication of articles authored by themselves in the journal they edit, possibly resulting in public scandals \cite{sternberg}. {There are, of course, perfectly benign reasons why editors self-publish, i.e., get their original research published in the journal they edit.} However, until now, the extent of this behaviour has remained unknown.

Although many have attempted to quantitatively investigate the self-publication of editors~\cite{bovsnjak2011analysis, luty2009preferential, mani2013publish, rosing2014publication, youk2019and, zdenvek2018analysis, walters2015editorial}, several aspects remain missing form the literature. In particular, existing studies do not compare the phenomenon across disciplines, since they focus on a single discipline each.
The only exception is the work of Bo{\v{s}}njak et al.~\cite{bovsnjak2011analysis}, but their dataset is restricted to Croatian journals. Furthermore, they only report average outcomes taken over the entire dataset, without providing a breakdown of their findings into disciplines. Existing studies also leave several disciplines entirely unexplored, including Biology, Business, Chemistry, Engineering, Geology, Materials science, Mathematics, Philosophy, Physics, and Political science. Moreover, several important comparisons are largely missing from the literature. First, by comparing the editors of a journal to other scientists who publish in that same journal, it might be possible to detect signs of favoritism, since the publication rates of an average scientist can serve as a benchmark. Second, by comparing the publication patterns of the editors before and after the start of their editorship, one might be able to detect editors who, right after they assume their editorial role, experience a sharp increase in the rate at which they publish in the journal. {Third, by comparing editors to their colleagues---those who serve on the same editorial board---one would account for the culture of the journal in question, thereby detecting editors whose publication rate is high relative to the norm in that journal.}
These critical comparisons are absent from the literature, except for the works of Campanario \cite{campanario1996competition}, Walters \cite{walters2015editorial}, and Mani et al.~\cite{mani2013publish}. The former two compared editors to other scientists, but only considered a single sub-discipline each, namely educational psychology and library information science, respectively. The latter paper compared the publication pattern before and after one becomes an editor, but only considered one sub-discipline, namely urology. A summary of the studies that considered the self-publication patterns of editors can be found in Supplementary Table~1.

{Another important aspect of editorial boards is their gender composition.} While many studies have examined this aspect~\cite{dickersin1998there, kennedy2001women, amrein2011women, ioannidou2015under, topaz2016gender, khan2019more, salazar2021gender}, no comprehensive review of these studies has been published to date, making it challenging to compare the different findings. Motivated by this gap in the literature, we conducted a thorough review of all papers published on this topic, and produced a comprehensive list of 50 papers, along with the discipline(s), the year(s), and the number of journals considered in each study; see Supplementary Table~2. Despite all these studies, important aspects remain missing. In particular, none of them compare the gender gap across disciplines, as they only focus on one discipline each. Moreover, certain disciplines were never considered by any of these studies, namely Chemistry, Computer Science, Engineering, Geology, Material Science, Philosophy, Physics, and Sociology. The only exception is the work of Mauleon et al.~\cite{mauleon2013assessing} who compared the gender gap across a wide range of disciplines, but only considered Spanish journals. Another limitation in the literature is the lack of a comparison between the proportion of female editors in a given discipline and the proportion of female scientists who are research active in that discipline over time. Again, the only notable exception is the work of Mauleon et al.~\cite{mauleon2013assessing} who, as mentioned earlier, is restricted to Spanish journals. Such a comparison is critical, as it provides a discipline-specific, and year-specific, benchmark against which gender disparity can be measured.

So far, we argued that the literature on both {self-publication} and gender disparity among scientific editors lacks a longitudinal study that compares editors to other scientists in their respective disciplines. To fill this gap, we parsed more than 173,000 editorial pages from Elsevier---a publisher behind one fifth of global research output, garnering one quarter of citations worldwide \cite{Elsevier_2020}. This enabled us to extract information pertaining to 103,000 editors, including their affiliations, their disciplines, the names of the journals they edited, and the years during which they served as editors. Collectively, these editors served on 85,000 issues of 1,167 Elsevier journals spanning 15 disciplines and multiple decades;
{see Supplementary Note~1 for details on how scientists' disciplines were inferred, and Supplementary Table~3 for the distribution of editors across disciplines.} Furthermore, following other studies in the literature \cite{topaz2016gender, alshebli2018preeminence, jadidi2018gender, holman2018gender, huang2020historical}, we used a state-of-the-art classifier (Methods), allowing us to identify the gender of 81,000 editors and 4,700 editors-in-chief with high confidence;
see Supplementary Table~4 for the distribution of those editors across disciplines.
Finally, to retrieve the publication record of any given editor, we use Microsoft Academic Graph (MAG)---a dataset of 220 million publications and 240 million scientists \cite{sinha2015overview, wang2019review}, which has been widely used in the Science of Science literature \cite{alshebli2018preeminence, huang2020historical, benson2018simplicial, frank2019evolution, murphy2020open, yang2020estimating, peng2021neural, gomez2022}. More specifically, we matched the editors in our dataset to the scientists in MAG based on their names, affiliations, and disciplines, thereby identifying the publication records of 20,000 editors and 1,600 editors-in-chief who had a unique match in MAG; see Methods for more details on the data collection, and Supplementary Table~5 for the distribution of those editors across disciplines.
The resulting datasets offer a unique opportunity to address the aforementioned shortcomings in the literature, and analyze editorial patterns at an unprecedented scale.

\section*{Results}
We start our analysis by exploring the characteristics of editors upon the start of their editorship and comparing them {to the average scientist}. To this end, let $(e,j)$ denote an editor-journal pair such that editor $e$ served on journal $j$. Moreover, let year$^{(e,j)}_1$ be the first year of the editorship, and let year$^{(e,j)}_0$ be the year that precedes it.
In our exploratory analysis, for every $(e,j)$, we randomly select {a scientist} whose {discipline and} academic birth year---the year when their first paper was published---matches that of $e$. Then, we compare {this scientist} to $e$ in terms of citation count, paper count, h-index, collaborator count, and affiliation rank. Note that the attributes of $e$ are measured in year$^{(e,j)}_0$, i.e., before the editorship has even started, implying that the measurements are not affected by $e$ becoming an editor of $j$. The scientist being compared to $e$ also have their attributes measured in year$^{(e,j)}_0$, implying that they have the same career length as $e$ at the time the measurements were taken. Finally, it should be noted that those scientists may themselves include editors of different publishing houses, as would be expected from the average scientist in MAG whom those {scientists} are meant to represent.

\begin{figure}[H]
\centering
\includegraphics[width=\textwidth]{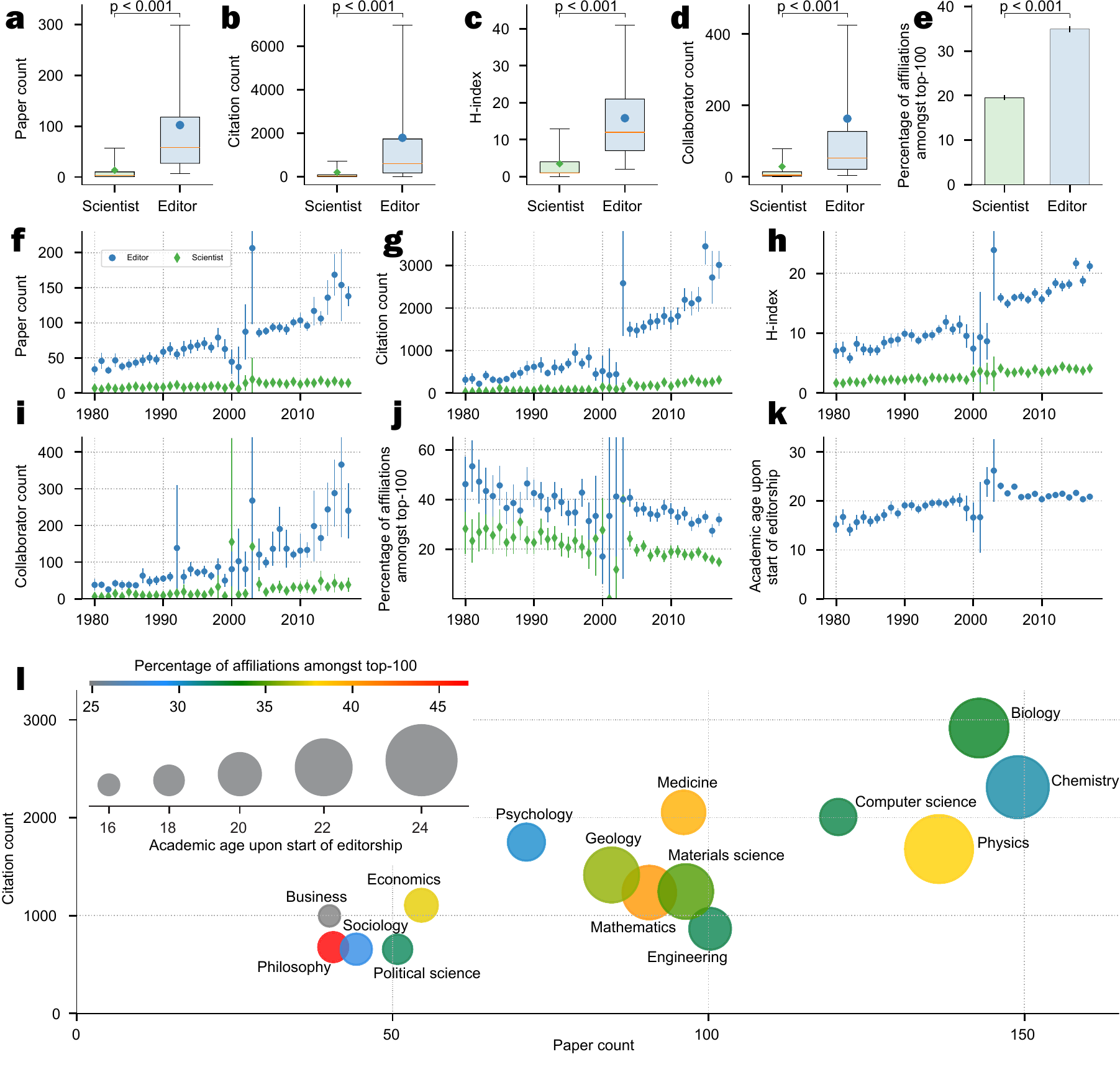}
\caption{\small{
\textbf{Editors' characteristics upon the start of editorship.} Each editor ($n=19,064$) is compared to a randomly selected scientist whose discipline and first year of publication matches that of the editor; descriptive statistics are measured at the year preceding the start of the editorship, with error bars representing the 95\% confidence intervals.
\textbf{a}--\textbf{e}, Comparing editors to scientists in terms of paper count, citation count, h-index, collaborator count, and percentage of those whose affiliation ranks among the top 100; { circles and diamonds represent the population mean of editors and scientists, respectively; the boxes extend from the lower to upper quartile values of the data, with a line at the median; whiskers extend until the 5-th and the 95-th percentile.}
\textbf{f}--\textbf{j}, Comparing editors to scientists over time in terms of paper count, citation count, h-index, collaborator count, and percentage of those whose affiliation ranks among the top 100.
\textbf{k}, For each year, the mean academic age of editors upon the start of their editorship.
\textbf{g}, Editors' paper count (x-axis), editors' citation count (y-axis), editors' academic age (circle size), and percentage of editors whose affiliation ranks among the top 100 (circle color) across disciplines; the differences in the circle sizes are exaggerated to improve visibility.
}}
\label{fig:descriptive}
\end{figure}

The results of this analysis are summarized in Figure~\ref{fig:descriptive}. As can be seen in Figures~\ref{fig:descriptive}a to \ref{fig:descriptive}c, an average editor has published 100 papers, has accumulated 1,800 citations, and has acquired an h-index of 16 by the start of the editorship. Compared to {an average scientist with the same academic age}, an editor tends to have {eight} times more papers, nine times more citations, and five times greater h-index. Note that these results disregard editorials; see Methods for more details on how editorials were identified. As for the number of collaborators, Figure~\ref{fig:descriptive}d shows that an editor has on average 160 at the start of the editorship, while {the average scientist} has about {30}. In terms of affiliation, Figure~\ref{fig:descriptive}e shows that 35\% of editors are affiliated with a top-ranked institution---one that is ranked amongst the top 100 according to the Academic Ranking of World Universities~\cite{UniversityRanking}---compared to just {20}\% for scientists. {Supplementary Figure~1 shows the distribution of the data in Figures~\ref{fig:descriptive}a to \ref{fig:descriptive}d}, while Supplementary Figures~2 to 6 show a breakdown of Figures~\ref{fig:descriptive}a to \ref{fig:descriptive}e over disciplines. {Moreover, instead of sampling a single scientist for each editor, we repeat the analysis twice, where 50 and 200 scientists are sampled with replacement for each editor; see Supplementary Figures 7 and 8. As can be seen, the results remain broadly unchanged.}

{
Next, we analyze how the characteristics of editors upon the start of their editorship have changed over the past four decades; see Figures~\ref{fig:descriptive}f to \ref{fig:descriptive}k.
More specifically, for any given year $y\in[1980,2017]$, we consider every editor-journal pair, $(e,j)$, such that year$^{(e,j)}_0 = y$, and measure the characteristics of $e$ at the year $y$; the average value is depicted in the figures as the blue circle at year $y$. Then, for every such $(e, j)$, {we also measure the characteristics of their matched scientist at the year $y$}; the average value is depicted as the green diamond at that year.} As shown in these figures, the expected number of citations that an editor has accumulated by the start of their editorship has increased tenfold over the past decades (from 300 in 1980 to 3,000 in 2017), the number of accumulated papers has more than quadrupled (from 33 to 138), the h-index has tripled (from 7 to 21), the number of collaborators has sixfold (from 40 to 240), while the percentage of those affiliated with top-ranked institutions has decreased (from {46\% to 32\%}). Next, we examine the gap between editors and scientists over the past decades. Comparing 1980 to 2017, we find that the gap in productivity has more than quadrupled, the gap in impact has increased {nine}-fold, the gap in h-index has more than tripled, while the gap in collaborator count has increased {more than six-fold}. As for the percentage of those affiliated with a top-ranked institution,
{it has decreased over the years for both editors and scientists at about the same rate, suggesting that this trend is not related to changes in the way editors are recruited, but rather to changes in the global demographics in academia.} {Again, these results remain unchanged when sampling 50 and 200 scientists per editor.} Finally, looking at the academic age of the editors upon the start of their editorship, we find that it increased from 15 years in 1980 to about 20 years in 2017; see Figure~\ref{fig:descriptive}k. {These findings suggest that, when it comes to assuming an editorial role, being impactful, productive, connected, and experienced seem to matter more than being affiliated with a top-ranked institution.}
Note that in Figures~\ref{fig:descriptive}f to~\ref{fig:descriptive}k an anomaly can be seen around the years 1998--2003. Upon inquiry, Elsevier clarified that this anomaly is an artifact of an incomplete capture of all articles during the first years of their transition from print to online.

{Having analyzed how different characteristics of editors change over time, we now compare those characteristics across disciplines.} Figure~\ref{fig:descriptive}{l} depicts the number of citations and papers that an editor has accumulated, as well as their affiliation rank and academic age upon the start of the editorship. We find that Biology recruits the most highly cited editors, with 2,900 citations on average, while Chemistry recruits the most productive editors, with an average of 149 papers. In contrast, impact seems to matter the least when recruiting editors in Philosophy, Sociology and Political Science, while productivity seems to matter the least when recruiting editors in Business and Philosophy. As for academic age, we find that Business recruits the youngest editors, with 16 years of experience on average, while Physics recruits the eldest, with 24 years of experience. We calculated the average academic age of editors across disciplines, and found it to be just over 20 years. Finally, in all disciplines, the percentage of editors affiliated with a top-ranked institution ranges from 25\% to 47\%, with Philosophy having the greatest percentage.

\begin{figure}[H]
\centering
\includegraphics[width=\textwidth]{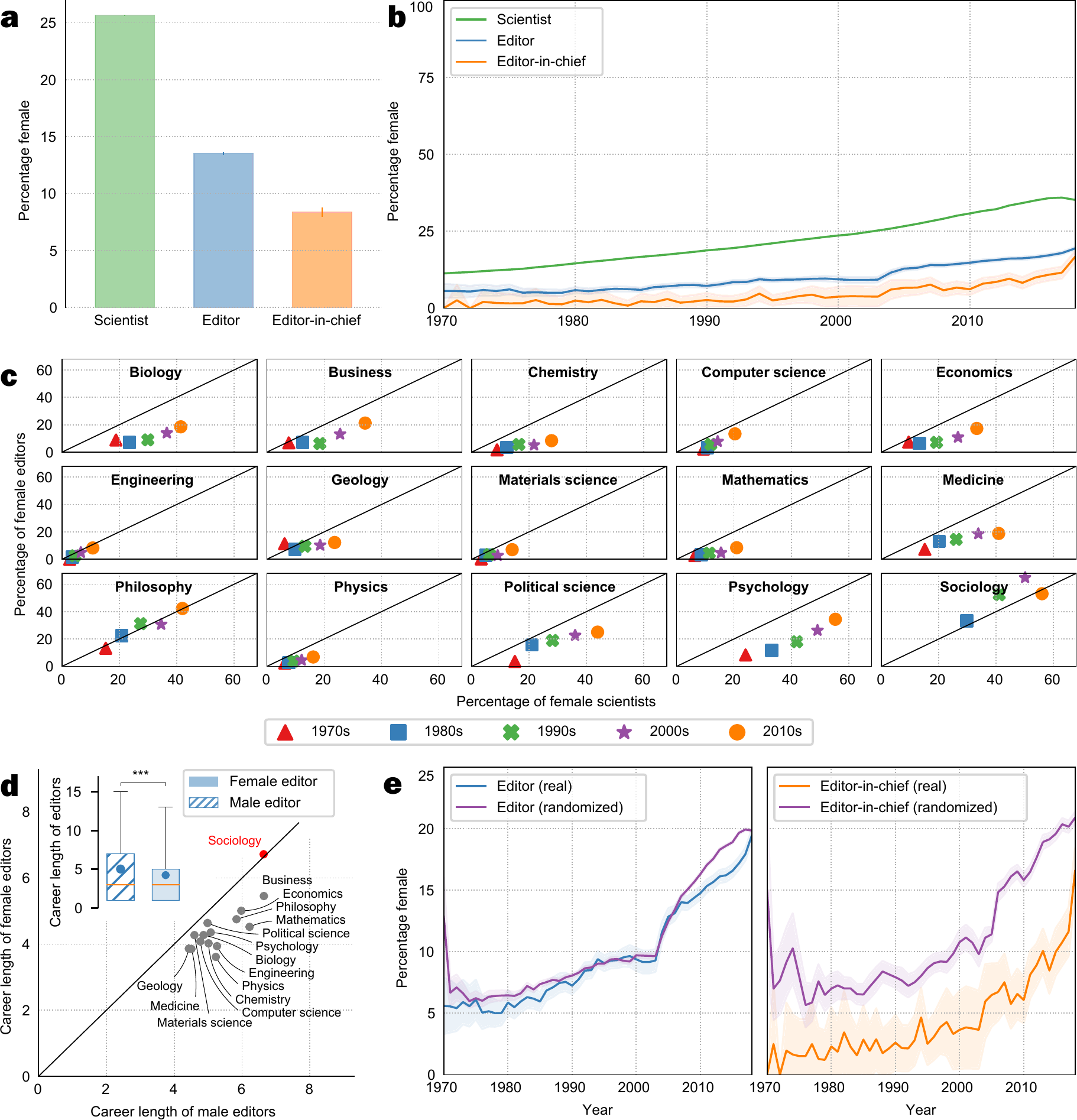}
\caption{\small{
\textbf{Gender disparity in editorship.}
\textbf{a}, Percentage of women among scientists ($n=42,831,834$), editors ($n=80,776$), and editors-in-chief ($n=4,692$).
\textbf{b}, Percentage of women among scientists, editors, and editors-in-chief over time.
\textbf{c}, Percentage of female editors against percentage of female scientists across disciplines in the 1970s (triangle), 1980s (square), 1990s (cross), 2000s (star), and 2010s (circle).
\textbf{d}, Average editorial career length of female vs.~male editors overall (inset) and across disciplines (scatter plot); red highlights the discipline in which the career length of female editors is greater than that of male editors.
\textbf{e}, Percentage of women among editors and editors-in-chief in real vs.~randomized data over time.
Error bars and shaded regions represent 95\% confidence intervals.
}}
\label{fig:gender}
\end{figure}

{We conclude our exploratory analysis by studying the gender disparity among editors. Figure~\ref{fig:gender}a compares the proportion of female scientists with the proportion of female editors and editors-in-chief.} {Although women are already underrepresented {among scientists} (26\% of all {unique scientists} in MAG), they are even more underrepresented amongst editors and editors-in-chief (14\% and 8\%, respectively).} Next, we compare the percentage of female editors and editors-in-chief to the percentage of female scientists over the past five decades. As shown in Figure~\ref{fig:gender}b, although gender parity has been steadily increasing in science in general, the proportion of female editors has consistently remained around half that of female scientists over the past five decades. For example, in 2017, women represented 36\% of scientists, but only 18\% of editors; these proportions are extremely similar to those in 1970, when women represented 11.3\% of scientists and 5.7\% of editors. As for female editors-in-chief, their proportion has remained consistently smaller than that of female editors since 1970.

Let us now examine the gender disparity across disciplines. Figure~\ref{fig:gender}c depicts the proportion of female editors against that of female scientists across disciplines during the 1970s (depicted as a triangle), 1980s (square), 1990s (cross), 2000s (star), and 2010s (circle). Apart from Sociology, the proportion of female scientists in any given discipline has remained greater than the proportion of female editors in that discipline; see how the vast majority of shapes fall under the diagonal. Similar patterns are observed for editors-in-chief; see Supplementary Figure~9. To obtain a better understanding of this phenomenon, we analyzed the editorial career length of editors, i.e., the number of years during which they assume their role. The box plot in Figure~\ref{fig:gender}d compares the average editorial career length of female vs.\ male editors, while the scatter plot compares these quantities across disciplines. As can be seen, the editorial career length of male editors is greater than that of their female counterparts ($t_{80,774}$ = 15.02, $P < 0.001$, $\beta$ = 0.15, 95\% CI = 0.13 to 0.16); this holds across all disciplines except Sociology. Supplementary Figure~10 shows similar patterns for editors-in-chief ($t_{4,685}$ = 6.27, $P < 0.001$, $\beta$ = 0.28, 95\% CI = 0.22 to 0.34), with the editorial career length of men being greater than that of women in all disciplines except Engineering, Geology, and Material Science.

As we have shown thus far, women have been consistently underrepresented in scientific editorship across disciplines over the past decades. Let us now investigate whether this phenomenon can be explained by gender differences in productivity, impact, and publishing career lengths, or whether additional hidden factors are at play. To this end, we use a randomized baseline model whereby each editor (or editor-in-chief) is replaced with a randomly chosen scientist who may have a different gender but is identical in terms of discipline and academic age, and similar in terms of productivity and impact (both binned into deciles). In this model, for any editor-journal pair, $(e,j)$, the randomly selected scientist who replaces $e$ serves as an editor (or editor-in-chief) on $j$ for the same number of years as $e$. Such a null model simulates a world where the editors in each discipline are recruited solely based on their experience and research output while completely disregarding their gender. We generated 50 such worlds and computed the average percentage of female editors and editors-in-chief therein. It should be noted that such analysis cannot be done using any of the datasets previously considered in the literature, as it requires the publication records of not only the editors but also all research-active scientists in any given discipline. The results of this analysis are depicted in Figure~\ref{fig:gender}e. As can be seen in the left panel, the representation of women among editors in a randomized world exhibits similar trends to those observed in the real world. This suggests that the gender gap among editors can be explained by the lack of women with sufficiently long publishing careers---a phenomenon that arguably explains the gender differences in productivity and impact in academia \cite{huang2020historical}. In contrast, looking at the right panel of Figure~\ref{fig:gender}e, we find a clear and persistent gap between the real and counterfactual worlds in terms of the proportion of female editors-in-chief. This suggests that factors other than career length, productivity, and impact may be at play, and these factors seem to persist over the past five decades.

Having analyzed the gender disparity in editorial boards, we now shift our attention to another interesting aspect of editorship---the fact that some editors publish original research in the journal they edit.
%
Our goal is to study whether becoming an editor of a journal is associated with an increase in the rate at which one publishes in that journal. To this end, it makes sense to start by analyzing the \textit{change in the publication rate} before vs.\ after assuming the editorial role. In this context, we focus on two intuitive and seemingly-extreme groups of editors, whose publication rates \textit{double} and \textit{triple}, after the start of their editorship; let us denoted these as $E_2$ and $E_3$, respectively.
More formally, for every $(e,j)$, let $\mathit{before}^{(e,j)}$ and $\mathit{after}^{(e,j)}$ denote the average number of papers per annum that $e$ publishes in $j$ during the five-year periods before and after year$^{(e,j)}_0$, respectively; see Methods for definitions.
Then, $E_2$ and $E_3$ are defined as follows:
{
\begin{subequations}
\begin{align}
\label{eqn:questionable}
E_2 &= \left\{ (e,j) : \mathit{after}^{(e,j)} \geq \max \Big(2 \times \mathit{before}^{(e,j)}, 1\Big)  \right\}\\
\label{eqn:suspicious}
E_3 &= \left\{ (e,j) : \mathit{after}^{(e,j)} \geq \max \Big(3 \times \mathit{before}^{(e,j)}, 2\Big)  \right\}
\end{align}
\end{subequations}
}
\noindent Note that the editors in $E_2$ not only double the rate at which they publish in $j$, but also publish at least one paper in $j$ annually after the start of the editorship. Similarly, editors in $E_3$ not only triple their rate, but also publish at least two papers in $j$ annually. These intuitive definitions imply that $E_2\subseteq E_3$. Editors who do not belong to either are referred to as $E_1$. Since our definitions require the publication patterns of editors during the five-year period following year$^{(e,j)}_0$, and since the publication records that we extracted from MAG do not go beyond 2018, we restrict this analysis to the 12,995 editor-journal pairs for which year$^{(e,j)}_0 \leq \textnormal{2013}$.

We will analyze each type of editor---{$E_1$, $E_2$, and $E_3$}---separately. Of course, an alternative approach would be to analyze the average trend of all editors combined. However, such analysis would be misleading if there are editors who drastically increase the rate at which they publish in their own journal soon after the start of their editorship. Such outliers may significantly increase the average rate overall, thereby giving the false impression that increasing one's publication rate in their own journal is common practice, when in reality only a few do so at an extremely high rate.
One may also argue that excessive publishing may be detected by examining the total number of papers that $e$ publishes in $j$, without comparing the rate before and after year$^{(e,j)}_0$. However, such an approach would flag any editor $e$ who was already publishing at a high rate before year$^{(e,j)}_0$. In contrast, our definition is more conservative, focusing only on those whose publication rate doubles (for {$E_2$}) or even triples (for {$E_3$}) after year$^{(e,j)}_0$. Finally, it should be noted that the definitions of the three types are meant to be intuitive; arguably, scientists from any discipline would appreciate the significance of doubling and tripling the rate at which editors publish in their own journal.

\begin{figure}[H]
\centering
\includegraphics[width=\textwidth]{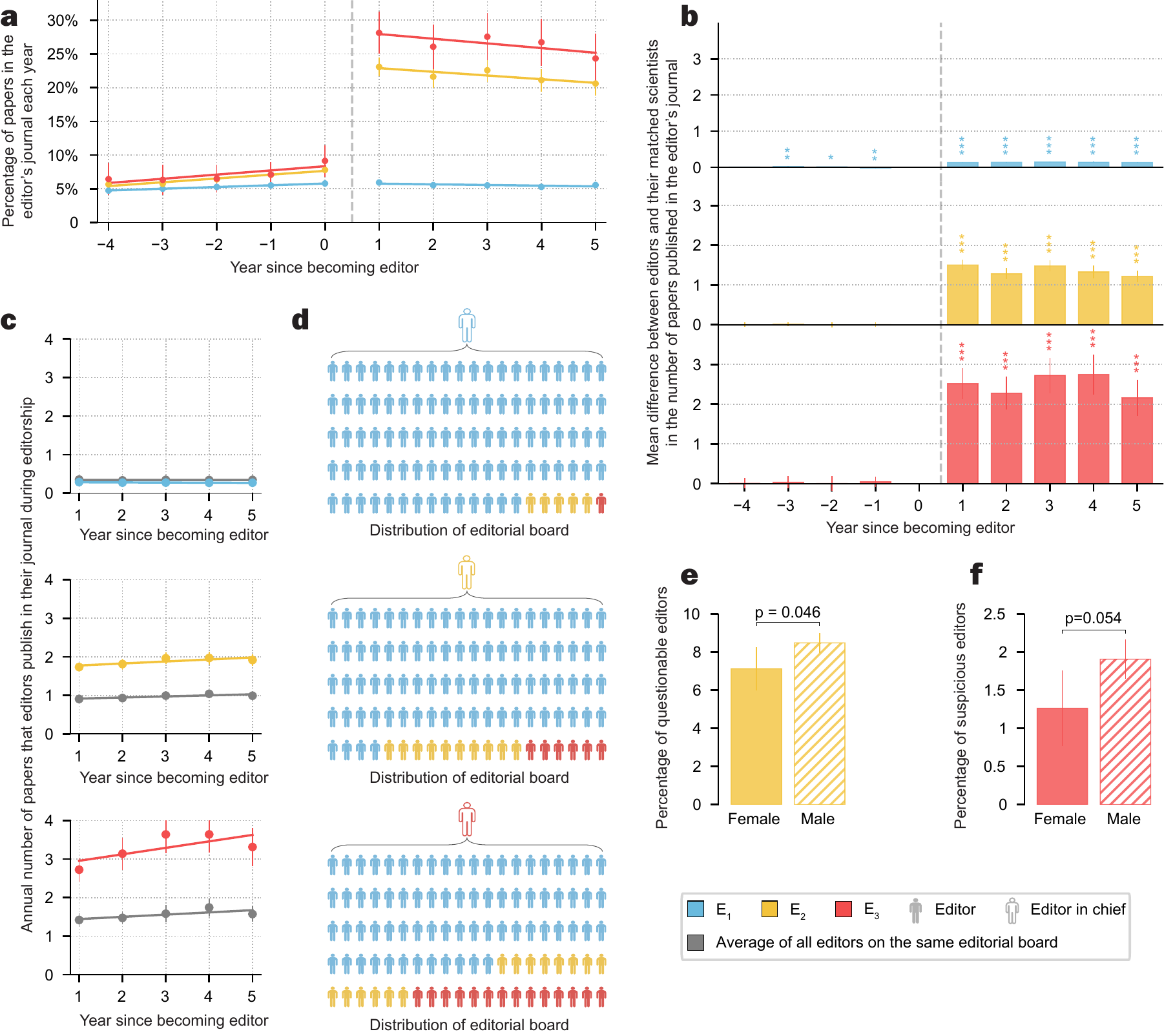}
\caption{\small{
\textbf{Analyzing $E_1$, $E_2$, and $E_3$ editors.}
Blue, yellow, and red correspond to $E_1$ ($n=11,920$), $E_2$ ($n=1,075)$, and $E_3$ ($n=235$) editors, respectively. Error bars correspond to 95\% confidence intervals.
{\textbf{a}, For every editor type, the average percentage of papers that $e$ publishes in $j$ per annum during the 5-year periods before and after the start of the editorship; the ``before'' period disregards the years before $e$'s academic birth year, while the ``after'' period disregards the years after $e$ leaves the editorial board of $j$.
\textbf{b}, Mean difference between $e$ and their matched scientists in terms of the number of papers published in $j$ before and after the start of the editorship;
the ``before'' period covers only the years following the academic birth years of both $e$ and their matched scientist, while the ``after'' period covers only the years during which $e$ is an editor of $j$ and the matched scientist is still research active.
\textbf{c}, Comparing $e$ to the average editor serving on the same editorial board.}
\textbf{d}, Distribution of editorial board given different types of editors-in-chief.
{
\textbf{e} and \textbf{f}, Percentage of {$E_2$ and $E_3$} editors, respectively, among male and female editors.
}
*$p<0.05$, **$p<0.01$, ***$p<0.001$.
}}
\label{fig:journal_outcomes}
\end{figure}

We start off by calculating the percentage of editors who fall under each type. We found that over 8\% of editors are in $E_2$ and about 2\% of editors are in $E_3$; these quantities persisted over the past decades as shown in Supplementary Figure~11.
One possible explanation is that, after becoming an editor, $e$'s productivity increased overall, implying that $e$ published more papers not only in $j$ but also outside of $j$. If this is indeed the case, we would expect the percentage of $e$'s papers that are published in $j$ to remain unchanged after the start of $e$'s editorship. To explore this possibility, Figure~\ref{fig:journal_outcomes}a depicts the percentage of $e$'s papers that are published in $j$ during the five years before, and after, year$^{(e,j)}_0$. As can be seen, this percentage triples for both $E_2$ and $E_3$ editors. In contrast, the percentage remains constant in the case of $E_1$ editors.

Another explanation could be that all scientists that are comparable to $E_2$ and $E_3$ editors experience an increase in productivity around that career stage. If this is the case, then the increase in $E_2$'s and $E_3$'s self-publication rate would be attributed to their career stage rather than the start of their editorship. To explore this possibility, for every editor-journal pair, $(e,j)$, we compare $e$ to randomly selected scientists who are not editors of $j$ but are similar to $e$ in terms of gender, discipline, rank of first affiliation, and years during which they are research-active. Additionally, we ensure that $e$ and their matched scientists are similar in terms of the outcome measure---the number of papers they publish in $j$---up to year$^{(e,j)}_0$. As such, if the outcome measure of the two starts to diverge after year$^{(e,j)}_0$, it suggests that the divergence is related to $e$ becoming an editor of $j$.
For more details on the matching process, see Methods.
Note that the matched scientists may themselves be editors of other journals. As such, the outcome of this analysis reflects the difference between those who edit $j$ and those who do not, rather than the difference between editors and non-editors.
The results of this analysis are depicted in Figure~\ref{fig:journal_outcomes}b.
In any given year, the bar height represents the difference between editors and their matched scientists in terms of the outcome measure. Apart from very few exceptions, no significant differences are observed before the start of the editorship, regardless of the editor's type. However, a significant difference ($p<$~0.001) is observed across all the years that follow. Notably, the magnitude of this difference is minuscule in the case of {$E_1$} editors, since they publish in their journal an average of only 0.1 additional papers per annum compared to their matched scientists. In contrast, those in {$E_2$} and {$E_3$} publish about 1.5 and 2.5 additional papers, respectively. The fact that the increase in {the number of papers published in one's own journal} happens right after year$^{(e,j)}_0$ suggests that it is strongly linked to them becoming an editor.

A third explanation for the observed increase in self-publication rate would be the journal's culture, whereby editors are expected to contribute papers as part of their editorial duties.
To determine whether the behavior of {$E_2$ and $E_3$} editors can be entirely explained by such {journal-specific factors}, we compare their publication rate to that of the average editor serving on the same editorial board; see Figure~\ref{fig:journal_outcomes}c. Indeed, different types of editors tend to serve on journals that differ greatly {in terms of publication rate}. More specifically, for journals on which an {$E_1$} editor serves, members of the editorial board publish in the journal an average of 0.3 papers per annum. In contrast, for journals on which {$E_2$ and $E_3$} editors serve, the rate is about 1 and 1.5, respectively. However, even when accounting for these differences, we still find evidence that those in {$E_2$ and $E_3$} publish excessively, since they publish at about twice the rate of their average colleague, while {$E_1$} editors publish at about the same rate.

Next, we shift our attention to the editors-in-chief. As can be seen in Figure~\ref{fig:journal_outcomes}d, given a journal whose editor-in-chief is an {$E_1$} editor, 94\% of the editorial board are in {$E_1$} and only 1\% are in {$E_3$}. In contrast, given an editor-in-chief who is in {$E_2$ ($E_3$)}, the percentage of {$E_1$} editors in the editorial board drops to 84\% (72\%) while the percentage of those in {$E_3$} increases to 6\% (14\%). These results suggest that the type of the editor-in-chief is related to the composition of the editorial board.

Finally, let us compare male to female editors in terms of the rate at which they publish in their own journal. As shown in Figure~\ref{fig:journal_outcomes}e, 8.5\% of male editors {fall in $E_2$}, compared to 7.1\% of female editors (Fisher's exact, $p =$~0.046).
Similarly, as shown in Figure~\ref{fig:journal_outcomes}f, 1.9\% of male editors {fall in $E_3$}, compared to 1.3\% for female editors (Fisher's exact, $p =$~0.054). Put differently, male editors engage in self-publishing more than female editors.

{Having established that $E_2$ and $E_3$ significantly increase their publication rate in their own journals after assuming their editorial roles, we conclude the current analysis by looking at the citation rate of such papers.} To this end, for every paper $p$ that an editor $e$ published in their journal $j$, we {measure} the number of citations it receives within the 5-year period post publication, excluding any citations received from other papers published in $j$; this measure is denoted by $c_5^{\textnormal{external}}$. Arguably, this exclusion {yields a measure of impact that is more robust against citation manipulation}, since any citations that $p$ receives from $j$ may have been requested (directly or indirectly) by $e$, or may have been added (consciously or subconsciously) by the authors to please $e$. To allow each paper to accumulate citations for five years, we restrict our analysis to the papers published before 2014. By plotting the average $c_5^{\textnormal{external}}$ of all papers that $e$ publishes in $j$ each year, we find very similar patterns across the three types of editors; see Supplementary Figure~12a. However, if we focus on the least-{cited} paper that $e$ publishes in $j$ each year, we find that {$E_3$} editors manage to get lower impact papers into their own journal when compared to those in {$E_2$}, who in turn get lower impact papers into their journals when compared to those in {$E_1$}; see Supplementary Figure~12b.

\begin{figure}[H]
\centering
\includegraphics[width=\textwidth]{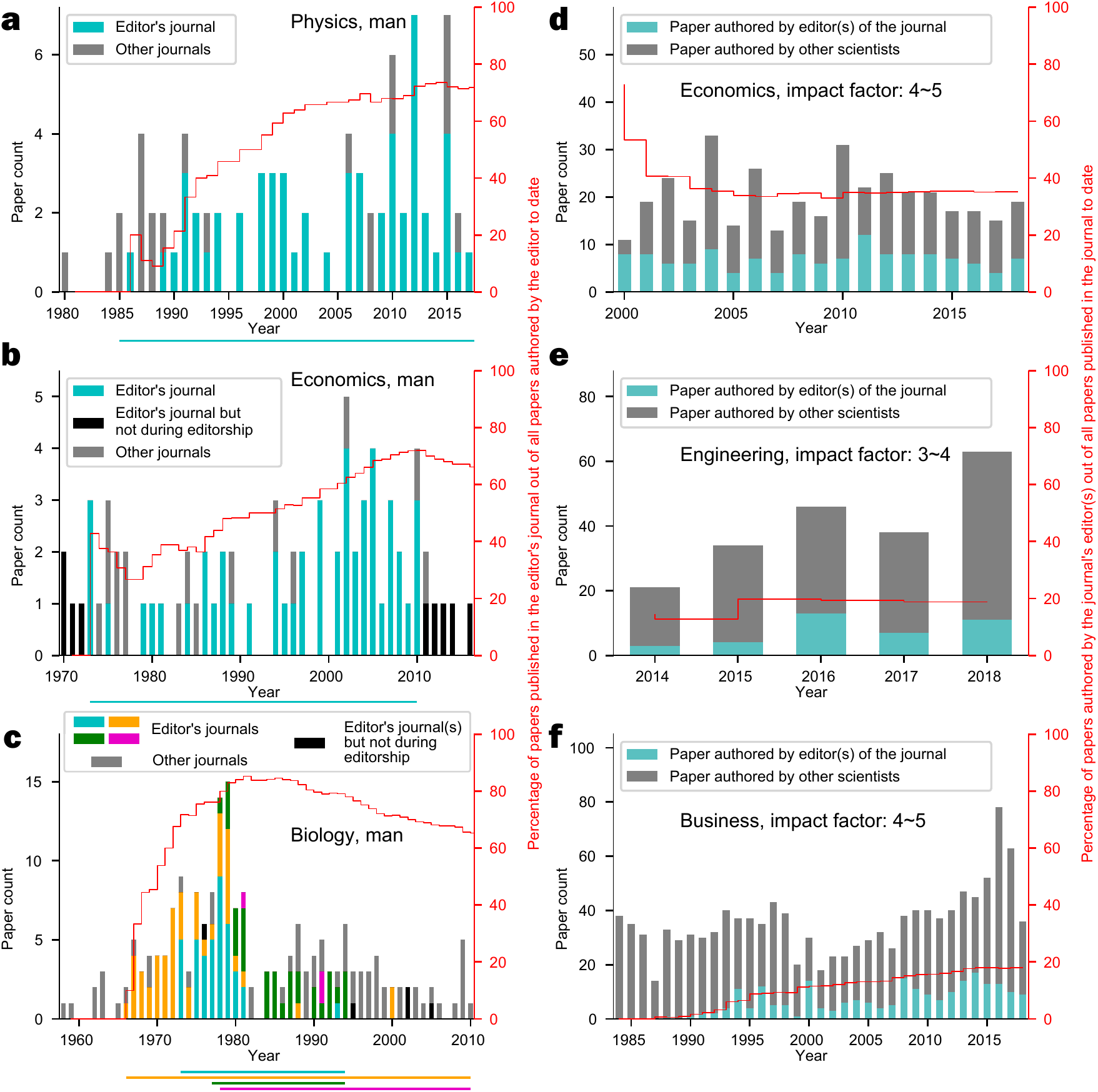}
\caption{\small{
\textbf{Extreme editors and extreme editorial boards.}
\textbf{a}--\textbf{c}, Out of all editors who publish at least 30 papers throughout their careers, the subfigures correspond to the three with the highest proportion of their papers published in the journal(s) they are editing. {For each of these editors, the corresponding subfigure specifies their gender, their discipline, the number of papers they publish each year, and how many of those papers are published in the editor's journal(s).} The horizontal line(s) underneath the plot represent the span of the editorship(s). For any given year, $y$, the red line (and the right y-axis) represents the proportion of papers published in the editor's journal(s) from the beginning of their scientific career until year $y$.
\textbf{d}--\textbf{f}, Out of all journals that have at least 30 papers by 2017, the subfigures depict the three with the highest proportion of papers whose authors include an editor of the journal. {For each of these journals, the corresponding subfigure specifies its discipline and impact factor (we avoid reporting the exact impact factor to preserve the anonymity of the journal).} For any given year, $y$, the red line (and the right y-axis) represents the aforementioned proportion, starting from the journal's launching year until year $y$. }}
\label{fig:outlier}
\end{figure}

{To understand the limits of the above phenomenon, i.e., the extent to which editors can increase their publication rate in the journal while continuing to serve on its editorial board,} we identified the {15} editors who publish the highest percentage of papers in their journals during editorship.
For each of them, we plotted the number of papers published per year throughout their scientific careers, highlighting in different colours the proportion of the papers published in the journal(s) they were editing; {see Figures~\ref{fig:outlier}a to \ref{fig:outlier}c for the three most extreme editors, and Supplementary Figure~13 for the remaining twelve}. In these figures, for every year, $y$, the red line (corresponding to the right y-axis) tracks the percentage of papers that the editor published in their own journal(s) starting from the beginning of their scientific career until year $y$.
The trends for the three most extreme editors are alarming; out of all the papers that they published throughout their career, 72\%, 66\%, and 65\% were published in the journal(s) they were editing while serving as editors.
{ Furthermore, these editors continued to publish excessively in their own journal(s) for many years after the start of their editorship, suggesting that it is unlikely for these papers to have been submitted before the editors' appointment.}
By taking a closer look at the most extreme editor (Figure~\ref{fig:outlier}a), we find that they published 56 papers in their own journal throughout their career, but not a single one of those papers was published outside the years during which they served as editor. Moreover, they published seven papers in their own journal in a single year (2012), without publishing any other papers in different journals that year. As for the second most extreme one (Figure~\ref{fig:outlier}b), a stretch of seven years stands out between 2003 and 2009. During those years, the editor published 16 papers in their own journal, but not a single paper in any other journal. Put differently, 100\% of the editor's papers were published in their own journal during those seven years. Finally, let us discuss the third most extreme editor (Figure~\ref{fig:outlier}c), who edited four different journals throughout their career, depicted in blue, yellow, green and red. Their rate peaked in 1981, when 85.2\% of their publications (98 out of all 115 that they authored by that year) were published in journals they were editing.
Also noteworthy is the fact that they authored 29 papers in just two years (1978 and 1979), all of which were published in three of the journals they were editing. After these two years, they continued to act as editors for these journals for 15 additional years (until 1994). These cases demonstrate that even if an editor publishes excessively in their journal, they may continue to serve as editors for several years afterwards. {Similar trends were observed when considering the 15 (rather than the three) most extreme editors; see Supplementary Figure~13. It is worth mentioning that 14 out of those editors are male, suggesting that women are less likely to engage in such extreme behavior.}

Having analyzed extreme editors, let us now focus on extreme editorial boards. To this end, we consider the three journals with the highest percentage of papers authored by their editors; see Figures~\ref{fig:outlier}d to \ref{fig:outlier}f, which follow a similar layout compared to our previous analysis of extreme editors. As for the first journal (Figure~\ref{fig:outlier}d), one third (35\%) of the papers published in it have an editor among the authors. As for the second and third journals (Figures~\ref{fig:outlier}e and \ref{fig:outlier}f), one fifth (about 20\%) of the papers include authors who happen to be editors. These cases demonstrate that editorial board members can author a substantial share of the papers published in the journal, and continue to do so for several decades.

\section*{Discussion}

Editorial boards of scientific journals are instrumental in promoting a fair and inclusive science~\cite{inclusive_academy}. Authors have argued for a deep connection between these aspects~\cite{corruption_marginalization}, but our understanding of them has also been hampered by a corresponding lack of longitudinal cross-disciplinary studies---a gap that was filled in our study. In particular, our unique dataset allowed us to study two fundamental aspects of editorship. The first is the gender distribution of editorial board members in any given discipline, and the degree to which it is representative of the gender distribution of scientists in that discipline. The second is {the rate at which editors publish in their own journal, and whether this rate increase soon after the editorship starts.}

Starting with the gender distribution, we found that only one in seven editors, and one in twelve editors-in-chief, is female. By examining the distribution over the past five decades, we found that the proportion of female editors persisted at about half that of female scientists, and the proportion of female editors-in-chief has consistently been smaller than that. We then examined the distribution across disciplines and found that women have been consistently underrepresented among editors and editors-in-chief in every scientific discipline other than Sociology. Moreover, compared to their male counterparts, female editors spend fewer years serving as editors and editors-in-chief in nearly all disciplines. We further investigated whether the observed gender disparity can be explained by the lack of women with sufficiently long publishing careers. The outcome suggests that this is indeed the case for female editors, but not editors-in-chief, implying that even if the difference in attrition rates between men and women are eliminated---a goal far from achieved to date---women are likely to remain underrepresented among editors-in-chief.

Moving on to {the self-publishing rate of editors}, we analyzed the years before and after the start of the editorship to detect any changes in the rate at which editors publish in their own journals. Here, we focused on two types---which we refer to as {$E_2$ and $E_3$}---who, after the start of their editorship, increase their self-publication rates by a factor of 2 and 3, respectively. We find that more than 8\% of editors {belong to $E_2$}, and about 2\% {belong to $E_3$}.
To exclude the effect of gender, discipline, affiliation, career stage, and prior productivity on the above increase in publication rate, we conducted a matching experiment that controls for these factors.
Furthermore, to rule out the possibility that this increase is entirely driven by journal culture, we compared {$E_2$ and $E_3$} editors to other members of the same editorial board.
The outcome of this analysis suggests that the increase in the self-publication rates of {$E_2$ and $E_3$} editors is indeed linked to them assuming their editorial roles.

Considering the composition of the editorial board, we find the proportion of {$E_2$ and $E_3$} members to be greater when the editor-in-chief happens to be {in $E_2$}, and even more so when they are {in $E_3$}, suggesting that the editor-in-chief's type is related to the editorial board composition. As far as gender is concerned, we find that female editors {experience a smaller increase in their self-publication rate} compared to male editors. Lastly, to understand the limits of the above behaviour, we examined the 15 editors with the most extreme behaviour, revealing that it is indeed possible for an editor to publish over 70\% of all their papers in the journal they are editing, and yet continue to assume their editorial duties for several decades.

Our findings contribute to a larger picture of academic publishing as a system with susceptibility to exploitation and inertia to increased openness~\cite{unesco}. For example, other studies have highlighted the prevalent manipulation of citations and authorship to get ahead in the academic reward system~\cite{wilhite2012coercive,fong2017authorship}, the bias exhibited by editors when handling papers that cite them~\cite{petersen2019megajournal}, and the emergence of journals where paying an open-access fee also guarantees a lenient peer-review~\cite{Bohannon60}. {Here, we showed that a considerable percentage of editors experience a sharp increases in self-publication rate after assuming their editorial roles, which may raise suspicion, as an external observer cannot rule out the possibility that the editor's papers were evaluated favorably.}
Our study also contributes to the literature advocating a more inclusive editorial board in specific, and a more inclusive scientific community in general \cite{inclusive_academy}. More importantly, gender disparity is often understood as a gap in terms of career length \cite{huang2020historical}, authorship \cite{lariviere2013bibliometrics}, and accumulation of citations \cite{lariviere2013bibliometrics, caplar2017quantitative}. We showed that, at least for editors-in-chief, gender disparity goes beyond what is predicted by these numbers.

{
There is more to the story of scientific publishing than statistics. Behind the numbers, some editors stand up for a fair selection of papers and actively recruit board members from underrepresented groups, while others exploit their power to benefit their careers. Editors are humans. Our expectations of human behavior in not perfectly transparent institutions determines the narrative: Is it remarkable that the vast majority of editors withstand the temptations of self-promotion? Or is it striking that more than 8\% percent {double their self-publication rate?}
Should we be satisfied with the increasing proportion of female editors over the past decades? Or should we be concerned that, despite all efforts to promote gender equality, women are still underrepresented in nearly all disciplines?
Either way, the culture of scientific editorship is not as fair and inclusive as it could be, and {we hope our study constitute a step towards achieve that goal}.
}

\section*{Methods}

\subsection*{Data}
\subsubsection*{Data Collection}
Elsevier publishes 4,289 different journals, all of which are listed on ScienceDirect---a website operated by the publisher~\cite{elsevierJournals}. Each journal curates some or all of its past issues, and all of the articles that appeared in every curated issue. In addition to research articles, many journals list their editors in the Editorial Board page, which can be found in the first volume of each issue. These pages, which constitute the primary source of our editor-related data, were retrieved using the Elsevier Article Retrieval API~\cite{elsevierAPI}. In total, we collected 173,434 editorial board pages from 1,893 different journals. From these pages, we were able to extract the following information about each journal: title, issue, volume, discipline, publication date, editors' names, editors' affiliations, and whether or not any given editor is an editor-in-chief. To retrieve the publication records of these editors, we paired them with scientists from the Microsoft Academic Graph (MAG) dataset. In particular, an editor in Elsevier and a scientist in MAG are considered to be the same person if, and only if, they uniquely share the same name and affiliation. {For details on how name disambiguation problem is addressed in this study, see Supplementary Note~2.}

{Editorials were then excluded from the publication record of each editor, to ensure that it consists of scientific papers. To this end, we queried ScienceDirect to identify the type of each publication in Elsevier, and excluded over 13,000 publications falling under the following types: Book review, Conference info, Editorial, Encyclopedia, Erratum, News, Practice guideline, and Product review. This left us with about 168,000 publications (co-)authored by the 20,000 editors identified in MAG. Out of those publications, we randomly sampled 200 and manually verified that only 2 were in fact editorial pieces. Additionally, we manually examined all 231 publications (co-)authored by the 3 extreme editors considered in Figure 4, and again found that only 2 were editorial pieces. This analysis suggests that our approach of identifying and excluding editorial pieces, while not perfect, is highly accurate.}

\subsubsection*{Gender Identification}\label{SM:GenderIdentification}
Several gender classifiers have been proposed to date \cite{lariviere2013bibliometrics, wais2016gender,west2013role}. Following other studies in the literature \cite{topaz2016gender,alshebli2018preeminence,jadidi2018gender,holman2018gender,huang2020historical}, we use \emph{Genderize.io}, which has been shown to outperform other alternatives \cite{wais2016gender}. This classifier integrates publicly available census statistics to build a name database, mapping names to binary gender labels. In our gender-related analysis, we only considered scientists whose first names were classified with a confidence of 90\% or above.

{
\subsubsection*{Dataset Evaluation}
In our gender-based analysis, we considered a dataset consisting of the 81,000 editors whose gender has been identified by \emph{Genderize.io} with confidence of $\geq$ 90\%; let us denote this dataset by $E_{\text{gender}}$. On the other hand, when analyzing the publication patterns of editors, we considered a dataset consisting of the 20,000 editors who had a unique matching entry in MAG, denoted by $E_{\text{MAG}}$. These two datasets are likely to exhibit biases since they were not randomly sampled from all 103,000 Elsevier editors, denoted by $E_{\text{all}}$. In this section, we aim to understand how these biases could affect our main findings.

We start off by comparing $E_{\text{gender}}$ to $E_{\text{all}}$ in terms of the relative size of each discipline. We found that the two are highly correlated ($r=0.99$); see Supplementary Figure~14a. High correlation could also be seen when comparing $E_{\text{MAG}}$ to $E_{\text{all}}$ ($r = 0.79$); see Supplementary Figure~14b. Nevertheless, some disciplines are under-represented compared to $E_{\text{all}}$, while others are over-represented. As a result, the observed proportion of women in $E_{\text{gender}}$ may differ from that in $E_{\text{all}}$, and the observed proportion of {$E_2$ and $E_3$} editors in $E_{\text{MAG}}$ may also differ from that in $E_{\text{all}}$. As shown in Supplementary Figure~14a, Medicine is over-represented in $E_{\text{gender}}$, which could affect our gender-based findings, especially since Medicine amounts to more than a fifth of all editors in $E_{\text{gender}}$. For instance, if this discipline happens to have fewer female editors than average, then our estimation of the gender gap (which is based on $E_{\text{gender}}$) would be an overestimation of the overall gender gap (i.e., the one in $E_{\text{all}}$).
To estimate the gender gap in $E_{\text{all}}$, we multiplied the proportion of women in each discipline in $E_{\text{gender}}$ by the size of that discipline in $E_{\text{all}}$. Similarly, to estimate the percentage of {$E_2$ and $E_3$} editors in $E_{\text{all}}$, we multiplied their proportion in each discipline in $E_{\text{MAG}}$ by the size of that discipline in $E_{\text{all}}$. As a result, the proportion of female editors becomes 12.95\% (instead of the originally-estimated 13.53\%), the proportion of {$E_2$} editors becomes 10.29\% (instead of 8.27\%), and the proportion of {$E_3$} editors becomes 2.62\% (instead of 1.81\%). This suggests that the situation may be more grim than what was originally estimated, since the gender gap seems to be larger, and the questionable behavior of editors seems to be more widespread, once we adjust for differences in discipline size.

Next, we examine how representative $E_{\text{MAG}}$ is of $E_{\text{all}}$ in terms of the editors' publication patterns. Since there are no available datasets that provide the publication profiles of all 103,000 editors in $E_{\text{all}}$, a practical alternative is to compare $E_{\text{MAG}}$ to a random sample of $E_{\text{all}}$, after manually identifying the publication profile of each sampled editor. Unfortunately, this approach also comes with its own limitations, since many editors do not have an online presence, making it extremely challenging, if not impossible, to manually identify their publication profile. With these limitations in mind, we sampled 500 editors from $E_{\text{all}}$ and were able to manually identify the MAG entry of 264 of them using information available online; the set of those 264 editors is denoted by $E_{\text{manual}}$. Then, we compared $E_{\text{MAG}}$ to $E_{\text{manual}}$ in terms of the confounders examined earlier in Figure~\ref{fig:descriptive}, namely: paper count, citation count, collaborator count, h-index, academic age, and percentage of those affiliated with a top-100 institution. We found significant differences between $E_{\text{MAG}}$ and $E_{\text{manual}}$ in terms of {paper count, collaborator count, h-index, and percentage of editors whose affiliation is among the top 100}; see Supplementary Figure~15. These differences may lead to a biased estimation of the percentage of {each type of editors}. {As a sensitivity analysis,} we re-sampled the editors in $E_{\text{MAG}}$ so that the distributions of {all confounders} are similar to $E_{\text{manual}}$. After re-sampling, we found no significant difference between $E_{\text{MAG}}$ and $E_{\text{manual}}$ in terms of the percentage of {$E_1$ and $E_2$}. This analysis suggests that the differences in {the above confounders} do not bias the estimation of the percentage of {$E_2$ and $E_3$ editors}. {Note that the aforementioned re-sampling is only done in this sensitivity analysis, and does not affect any other result in our study.}

Finally, let us directly compare $E_{\text{manual}}$ to $E_{\text{gender}}$ in terms of the percentage of female editors, and compare $E_{\text{manual}}$ to $E_{\text{MAG}}$ in terms of the percentage of {$E_2$ and $E_3$} editors. Out of the editors in $E_{\text{manual}}$, 12.54\% were female (compared to 13.53\% in $E_{\text{gender}}$), 8.71\% were {$E_2$} editors (compared to 8.27\% in $E_{\text{MAG}}$), and 3.41\% were {$E_3$} editors (compared to 1.81\% in $E_{\text{MAG}}$); none of the differences were statistically significant as shown in Supplementary Figure~16.
}

{
\subsection*{Definitions }\label{SM:Definitions}

Let $E$ denote the set of editors, let $J$ denote the set of journals, and let $P \subseteq E\times J$ denote the set of editor-journal pairs $(e, j)$ for which $e$ served as an editor of $j$. For any editor-journal pair, $(e, j)$, the first (last) year of editorship is assumed to be the publication year of the first (last) issue of $j$ in which $e$ is mentioned as an editor. Moreover, the editorial career of $e$ (as an editor of $j$) is assumed to span the period between the first and last years of editorship (inclusive), implying that any gap years (if they exist) are included in our analysis. Similarly, the academic career of any scientist $s$ is assumed to span the period between the publication years of their first and last papers. As a result, the academic age of $s$ in any given year $y$ is $y - {\textnormal year}^s_{\text{first}} + 1$, where year$^s_{\text{first}}$ is the publication year of the first paper of $s$.

Finally, let us formally define the periods $\mathit{before}^{(e,j)}$ and $\mathit{after}^{(e,j)}$. To this end, let year$^{(e,j)}_1$ be the first year in which $e$ edits $j$, and let year$^{(e,j)}_0$ be the year that precedes it. Moreover, let year$^{(e, j)}_{\text{end}}$ denote the year in which $e$'s editorship of $j$ ends. Finally, for any two years, $a,b:a \leq b$, let $\mathit{avg}^{(e,j)}(a,b)$ denote the average number of papers per annum that $e$ publishes in $j$ during the years $[a,b]$. Then, the periods $\mathit{before}^{(e,j)}$ and $\mathit{after}^{(e,j)}$ are defined as follows:
$$
\mathit{before}^{(e,j)} =
    \left\{
        \begin{array}{ll}
            \mathit{avg}^{(e,j)}(\text{year}^e_{\text{first}},\text{year}^{(e,j)}_0) & \textnormal{if } \text{year}^{(e,j)}_0 - \text{year}^e_{\text{first}} < 5 \smallskip\\
            \mathit{avg}^{(e,j)}(\text{year}^{(e,j)}_{-4},\text{year}^{(e,j)}_0) & \textnormal{otherwise.}
        \end{array}
    \right.
$$
$$
\mathit{after}^{(e,j)} =
    \left\{
        \begin{array}{ll}
            \mathit{avg}^{(e,j)}(\text{year}^{(e,j)}_1, \text{year}^{(e,j)}_{\text{end}}) & \textnormal{if } \text{year}^{(e,j)}_{\text{end}} - \text{year}^{(e,j)}_0 < 5 \smallskip\\
            \mathit{avg}^{(e,j)}(\text{year}^{(e,j)}_1,\text{year}^{(e,j)}_5) & \textnormal{otherwise.}
        \end{array}
    \right.
$$

\noindent where year$^{(e,j)}_x: x\in \{-4,\ldots, 5\}$ is $x$ years away from year$^{(e,j)}_0$, implying that ${\text{year}}^{(e,j)}_{-4}$ and ${\text{year}}^{(e,j)}_{5}$ are 4 years before, and 5 years after, ${\text{year}}^{(e,j)}_0$, respectively. Intuitively, $\mathit{before}^{(e,j)}$ and $\mathit{after}^{(e,j)}$ are the 5-year periods before, and after, the start of $e$' editorship of $j$, respectively, with the only exception being that $\mathit{before}^{(e,j)}$ disregards the years before $e$'s academic birth year, while $\mathit{after}^{(e,j)}$ disregards the years after $e$ leaves the editorial board of $j$.
}

\subsection*{Matching Editors to Scientists}\label{SM:matching}

Given an editor-journal pair $(e, j)$, we match $e$ to a scientist
{$s$ who is not an editor of $j$}
based on a number of confounders, including the rate at which they publish in $j$. Ideally, the rate of $e$ and $s$ should be similar up to year$_0^{(e,j)}$ (this way, if their rate starts to diverge after year$^{(e,j)}_0$, it suggests that the divergence is related to $e$ becoming an editor of $j$). However, to increase the likelihood of finding a match for $e$, we do not require the rate of $s$ to match that of $e$ in year$_0^{(e,j)}$, but rather in a year $y$ such that $\vert \textnormal{year}_0^{(e,j)} - y\vert \leq 3$.
More specifically, we say that $e$ matches $s$ if all of the following conditions are met:

\begin{itemize}
\item $e$ and $s$ have the same discipline.

\item $e$ and $s$ have the same gender; for details on how gender is identified, see the subsection titled Gender Identification.

\item The rank of any first known affiliations of $e$ and $s$ fall in the same bin. Here, affiliations are ranked based on the 2019 \emph{Academic Ranking of World Universities} (also known as the ``Shanghai ranking''~\cite{UniversityRanking}), and are divided into the following bins: $[1,20]$; $[21, 50]$; $[51,100]$; $[101,300]$; $[301,600]$; $[601,999]$; $[1000,\infty]$.

\item The publication year of $e$'s first paper does not differ from that of $s$ by more than 3 years.

\item There exists a year, $y \in [\textnormal{year}_0^{(e,j)}-3, \textnormal{year}_0^{(e,j)}+3]$ such that:

\begin{itemize}
    \item The academic age of $e$ in year$_0^{(e,j)}$ does not differ from that of $s$ in $y$ by more than 10\%.

    \item The number of papers that $e$ published in $j$ in year$_0^{(e,j)}$ does not differ from that of $s$ in $y$ by more than 10\%.

    \item The total number of papers that $e$ published in $j$ up to year$_0^{(e,j)}$ does not differ from that of $s$ in $y$ by more than 10\%.
\end{itemize}
\end{itemize}

\section*{Data Availability}
Our editors' dataset was collected from Elsevier's ScienceDirect database. A formal agreement between us and Elservier mandates that data copied from the subscribed products cannot be provided to third parties in any substantial or systematic manner. However, for transparency reasons, we provide a sample set of 10 editors, which can be used to test the code for data collection and analysis, along with anonymized data for reproducing figures, all the while ensuring that our agreement with Elsevier is not breached. As for our publications' dataset, it can be retrieved from Microsoft Academic Graph's official website; a small subset of MAG that is sufficient to test our code is also provided. To retrieve the aforementioned datasets, visit \textcolor{blue}{\href{https://github.com/Michael98Liu/fair-and-inclusive-scientific-publishing/tree/main/data}{this link}}.

\section*{Code Availability}
The code used to collect and clean the editors' dataset, as well as the code used in the analysis and visualizations, are all freely available to download \textcolor{blue}{\href{https://github.com/Michael98Liu/fair-and-inclusive-scientific-publishing}{here}}.


\section*{Acknowledgement}
F.L. was supported by the New York University Abu Dhabi Global PhD Student Fellowship. P.H. acknowledges financial support from JSPS KAKENHI grant no. JP21H04595. The support and resources from the High Performance Computing Center at New York University Abu Dhabi are gratefully acknowledged. Finally, we would like to thank Aikaterini Kyriazidou and Christian Haefke, as well as the three anonymous reviewers, for their feedback and comments which improved the paper.

\section*{Author Contributions}
Conceptualization: BA, MC, TR;
Methodology: BA, FL, PH, TR;
Data collection and processing: BA, FL;
Visualization: BA, FL, PH, TR;
Supervision: BA, TR;
Writing: BA, FL, PH, TR.

\section*{Competing interests}
The authors declare no competing interests.

\end{document}